\documentclass{aa}

\usepackage{epsfig}

\begin{document}

%\input{psfig}
%   \thesaurus{06     % A\&A Section 6: Form. struct. and evolut. of stars
%              (03.11.1;  % Cosmogony,
%               16.06.1;  % Planets and satellites: general,
%               19.06.1;  % Solar system: general,
%               19.37.1;  % Stars: formation of,
%               19.53.1;  % Stars: oscillations of,
%               19.63.1)} % Stars: structure of.
%
   \title{Flares observed with {\em XMM-Newton} and the VLA }

   \author{Kester Smith\inst{1}\thanks{Present address: STScI, 3700 San Martin Drive, Baltimore MD 21218, USA}
          \and
	  Manuel G\"udel\inst{2}
          \and
          Marc Audard\inst{3}
          }

   \offprints{K. Smith}

   \institute{MPIfR, Auf dem H\"ugel 69, 53121 Bonn, Germany
	      \and
Paul Scherrer Institut, W\"{u}renlingen und Villigen,  
              CH-5232 Villigen PSI, Switzerland. 
	      \and
	      Columbia Astrophysics Laboratory, 
	      Columbia University, 
	      550 West 120th Street, 
	      New York, NY 10027, 
	      USA \\
              kester@mpifr-bonn.mpg.de  \\
              guedel@astro.phys.ethz.ch \\
              audard@astro.columbia.edu \\
             }

   \date{Received 22$^{\mathrm{nd}}$ September 2004, ; Accepted 16$^{\mathrm{th}}$ December 2004 }

\abstract{We present lightcurves obtained in X-ray by the {\em XMM-Newton} 
EPIC cameras and simultaneous radio lightcurves obtained with
the VLA for five active M-type flare stars. A number of
flare events were observed, 
and by comparing radio with X-ray data, we consider various
possible flare mechanisms. In cases where there seems to be a clear
correlation between radio and X-ray activity, we use an energy budget
argument to show that the heating which leads to the X-ray emission
could be due to the same particles emitting in the radio. In cases
where there is radio activity without corresponding X-ray activity, we
argue that the radio emission is likely to arise from coherent
processes involving comparatively few particles. In one case, we are
able to show from polarization of the radio emission 
that this is almost certainly the case.
Cases for which X-ray activity is seen without corresponding radio
activity are more difficult to explain. We suggest that the heating
particles may be accelerated to very high energy, and the resulting synchrotron radio
emission may be beamed in directions other than the line of sight.
}

\maketitle

%
%________________________________________________________________

\section{Introduction}

Convection and differential rotation in late-type stars combine to
drive a magnetic dynamo, leading to the formation of a corona. The
solar corona is readily observable due to the Sun's proximity. In the
case of other stars, magnetic activity may manifest itself through
various chromospheric emission lines, quiescent soft X-ray and
microwave emission, and the occurence of flaring behaviour.  Stars
with rapid rotation, either due to a young age or membership of
close binary systems, display much more extreme coronal activity than
the Sun. Numerous small flares are the leading candidate for the main
mechanism of coronal heating on the Sun (Krucker \& Benz 1998), and
there is growing evidence that the same mechanism could be responsible
for coronal heating in other, more magnetically active stars
(e.g. Audard et al. 2000,
Kashyap et al. 2002, G\"udel et al. 2002b, 
G\"udel et al. 2003, 
Audard et al. 2003).

Standard models of solar flares envisage the key process to be
acceleration of electrons and ions by magnetic reconnection (see e.g.
Benz 2002). The energy in the electron (or ion) beam then gives rise
to the flare emission. Three main emission mechanisms can be
identified.  First, radio gyrosynchrotron emission from the
magnetically confined electrons, which accounts for a small fraction
of the energy. Second, hard X-ray emission which is produced as the
electron beams collide with the chromosphere, and which also accounts
for a small portion of the energy.  Third, soft X-ray emission is
produced from evaporating chromospheric material heated to around
10$^7$K (Dennis 1988). This accounts usually for the bulk of the
emitted energy. There is, therefore, a physical relationship between
the radio gyrosynchrotron emission and the X-ray emission, which, on
the Sun, manifests itself in various observable effects for individual
flares. Temporal and spatial correlations between components of
individual flare events can be investigated to build up a detailed
picture of the flaring process. For flare stars, however, spatial
information is not available, and temporal resolution is heavily
compromised by the need for integration times sufficient to yield
acceptable signal-to-noise ratios.  We must therefore search for
coarser relationships between different emission types, which can
point to similarities with or differences to the solar models. An
indication that radio emission and soft X-ray emission are generally
linked comes from the relation between the quiescent radio emission
and X-ray luminosities, which holds over a wide range of flare
energies from different types of flaring object (G\"udel \& Benz
1993).  Temporal relations can also sometimes be observed. For
example, the Neupert effect, in which the soft X-ray flux is
proportional to the integrated radio flux, is evidence that heating of
and evaporation from the chromosphere is driven by the same population
of fast particles which gives rise to the radio emission (Neupert
1968, Dennis \& Zarro 1993). Examples of the Neupert effect in stellar
sources include G\"udel et al. (1996), for UV~Cet and recent
observations of $\sigma$ Geminorum (G\"udel et al. 2002a) and Proxima
Centauri (G\"udel et al. 2002b).

The new generation of X-ray satellites, offering greatly enhanced
sensitivity in the soft X-ray regime, makes it easier to pursue
simultaneous radio and X-ray monitoring programmes aimed at flare
stars, with a view to searching for temporal correlations between
radio and X-ray activity.  In this paper we present a collection of
time series observations for five active stars, all of which are M
dwarfs.  In each case, we have combined VLA radio observations with
simultaneous {\em XMM-Newton} EPIC observations, to search for any
relationship between the gyrosynchrotron radio and soft X-ray
emission.

\section{Target objects}

The five targets discussed in this paper are all active late-type
stars known to show strong coronal activity.  AD~Leo is an dM3.5e star
of apparent magnitude V=9.43 lying at a Hipparcos distance of
approximately 4.7~pc. It shows an apparent rotation period of 2.7 days
(Spiesman \& Hawley 1986) and v$\sin{i}$ of 5 kms$^{-1}$ (Vogt, Penrod
\& Soderblom 1983) which together suggest a high inclination between
the stellar rotation axis and the plane of the sky ($\sim$75$\deg$).  The
system is in fact binary, with a very low mass infrared companion at a
separation of 78~mas, or 0.366~AU (Balega, Bonneau \& Foy 1984). The
quiescent and flaring X-ray emission were studied by Favata et
al. (2000) who used a combination of {\em Einstein}, ROSAT and ASCA
data and drew conclusions about the nature of magnetic structures in
the flaring corona. Stepanov et al. (2001) observed an intense radio
flare from the star, which was interpreted as coherent emission from a
flare loop. G\"udel et al. (2003) studied the statistical flaring
properties of AD~Leo, fitting power laws to the flare energy
distribution to investigate the possibility of coronal heating being
caused by small flares.  Hawley et al. (1995, 2003) conducted a
multiwavelength study of flaring behaviour, combining observations
from ground based optical observatories, HST STIS spectroscopy, EUVE
data, as well as microwave observations from MERLIN.  They observed
several examples of a Neupert effect between the EUVE soft X-rays and
the U-band emission, which is a proxy for hard X-ray emission. Van den
Besselaar et al. (2003) presented {\em XMM-Newton}\ and Chandra spectra of AD
Leo (and also showed the {\em XMM-Newton}\ lightcurve presented here), and
derived various properties of the corona, including the possible
presence of an inverse FIP effect.

AU Mic is an active dM0e dwarf at a distance of 10pc with a
rotation period of 4.85 days (Vogt et al 1983). It forms a distance
and proper motion pair with AT~Mic, although the two are separated by
1.5$^{\circ}$ on the sky. Detailed modelling of the hydrogen
spectrum by Houdebine \& Doyle (1994) revealed many of the conditions
prevailing in the chromosphere and transition region.  Microflaring on
timescales of seconds was detected in {\em U}-band photometry by
Andrews (1989). X-ray variability was studied by Ambruster et
al. (1987). Kundu et al. (1987) observed AU Mic, and also AT Mic, at
microwave frequencies with the VLA.  A massive flare was observed by
EUVE on this object in July 1992, and is discussed by various authors,
including Cully et al. (1994) who modelled the slow decay phase as
being due to the ejection of a magnetically confined plasmoid, an
event similar to a solar coronal mass ejection. The time-varying UV
spectral lines during this event were discussed by Monsignori Fossi et
al. (1996) and also by Katsova et al. (1999). The quiescent UV
spectrum was studied by Pagano et al. (2000), who derived an emission
measure distribution and compared the conditions to those on the
Sun. FUV flare behaviour was studied by Robinson et al. (2001) using
HST spectra.
 
AT~Mic is a dM4.5e+dM4.5e binary with an apparent separation of
3$''$.5 at a distance of 10.2~pc.  UV and optical flaring behaviour
was described by Bromage et al. (1986).  and simultaneous optical,
infrared and microwave observations were made by Nelson et al. (1986).
The microwave spectrum was studied by Large et al. (1989) who
concluded that two components were present, variable emission from a
coherent process producing a falling spectrum below about 1~GHz, and a
probable gyrosynchrotron process producing a flat spectrum. Gunn et
al. (1994) saw a blue shifted component in the profile of the Ca II H
and K lines during a flare, which they interpreted as evidence of
chromospheric evaporation.  X-ray spectra of AT~Mic obtained by
{\em XMM-Newton}\ and Chandra were presented by Raassen et al (2003). This
paper also presented the EPIC pn lightcurve analysed here.

\begin{table*}[!t]
\caption{\label{observations} Summary of the target objects and
  observations log.  The final column lists the duration of the
  simultaneous observation, i.e. the length of time over which both
  X-ray and radio data were obtained. In the case of UV~Cet, this
  duration is identical for the 5~GHz and 8.3~GHz data }
\begin{center}
\begin{tabular}{|l|l|l|l|l|c|} \hline
Source    & Spectral Type   & d    & P   & Observing date & Simultaneous observation length  \\
          &                 & (pc) & (d) &                & (ks)       \\          
\hline
AD Leo    &  dM3.5e         & 4.7  & 2.7  &  14.05.2001   & 31.54  \\      
AU Mic    &  dM0e           & 10   & 4.85 &  13.10.2000   & 20.74 \\ 
AT Mic    &  dM4.5e+dM4.5e  & 10.2 &     &  15.10.2000    & 16.16  \\
UV Cet    &  dM5.5e+dM5.5e  & 2.7  &     &  07.07.2001    & 30.24 \\ 
YZ CMi    &  dM4.5e         & 5.9  &     &  09.10.2000    & 20.04 \\
\hline
\end{tabular}
\end{center}
\end{table*}

UV Cet is a dM5.5e+dM4.5e binary at a distance of 2.7~pc.  The
projected separation on the sky is about 2$''$. Radio observations
have shown significant differences in the behaviour of the two stars,
despite their similar basic properties.  The quiescent primary is much
weaker than the secondary, and often produces highly polarized flares,
suggesting a coherent emission mechanism. The secondary flares more
frequently, but produces only moderately polarized flares suggestive
of a gyrosynchrotron origin.  VLBA observations of UV Ceti B show that
the radio source is extended to about 4 -- 5 stellar radii. This is
thought to indicate trapping of accelerated particles in large coronal
loops. The primary was found to be pointlike in these observations.
The two components were recently distinguished in X-rays for the first
time by Audard et al. (2003).  They found that the quasi-steady X-ray
behaviour of the two components was similar, in contrast to the
quiescent radio emission. The apparent quiescent emission of component
B was probably composed of numerous small flares. The brighter radio
emission of UV~Cet~B may be due to more effective trapping of accelerated
particles in the large coronal loops of this object.

YZ CMi is a dM4.5e star at a distance of
approximately 5.9~pc.  UV observations by Robinson et al. (1999)
revealed flaring activity on a range of time and energy scales. VLBI
observations by Pestalozzi et al. (2000) resolved the corona at 3.6~cm
to extend $1.77 \times 10^{10} \pm 8.8 \times 10^9$~cm, or $0.7 \pm
0.3$~R$_{*}$  above the photosphere.

\section{Observations}

 The X-ray data are from the {\em XMM-Newton}\ guaranteed time
  programme.  The X-ray lightcurves were constructed using standard techniques with
{\em SAS},  and were smoothed by using a Savitzky-Golay
smoothing algorithm (as implemented in IDL). The smoothing employed
here uses in each case an 11-point 4th-degree kernel.  The
Savitzky-Golay filter provides an estimate of the first derivative of
the curve, which is of interest when considering possible correlated
X-ray and radio events. The X-ray hardness ratio was also measured. We
defined this as the hard count rate, in a band between 1.5 and
4.5~keV, divided by the soft count rate, in the band 0.3 to
1.5~keV. 

The VLA observations were scheduled to be simultaneous with the {\em
  XMM-Newton}\ X-ray observations. Most of the observations were
carried out in October 2000, with the VLA in C configuration.  All the
radio observations for AD~Leo, AU~Mic, AT~Mic and YZ~CMi were made at
6~cm. UV~Cet was observed at both 6~cm (5~GHz) and 3.6~cm
(8.3~GHz).  The low frequency band was chosen for most of the
  observations primarily because the flare gyrosynchrotron emission is
  expected to exhibit a falling spectrum, so that the detection
  likelihood for marginal events is maximised. The need to observe
  phase calibrators introduced gaps into the time coverage. Typically,
  the target would be observed for approximately three to four
  minutes, and then a calibrator would be observed for approximately
  two minutes.  The data were reduced using standard tasks in
  AIPS. The integration time was ten seconds, which is then the
  minimum available binning time in the data reduction. Lightcurves
  were produced by coherently averaging the real and imaginary parts
  of the visibility on a timescale of typically one scan, before
  finally determining the stellar flux. The radio data could be
  rebinned at higher resolution to closely study a few interesting
  events. 

The observations and basic data for the target objects are summarized
in Table~\ref{observations}.

\section{Overview of the analysis}
\label{overview}

We pay particular interest to time correlations between X-ray and
radio events.  
Approximately simultaneous X-ray and radio events suggest the
possibility of the Neupert effect. The correlation arises because the
gyrosynchrotron (radio) emission is proportional to the instantaneous
number of fast particles, whilst the slowly variable soft X-ray
emission is roughly proportional to the accumulated total
energy. Thus,
\begin{equation}
L_R(t) \propto \frac{d}{dt}L_X(t), 
\end{equation}
where $L_R$ and $L_X$ are the radio and soft X-ray luminosities.  It
is for this reason that the first derivative of the X-ray lightcurve
is of interest. If flaring behaviour in radio is correlated quite
closely with the gradient of the X-ray lightcurve, this then provides
an {\em a priori} implication that the Neupert effect may be occuring
in a given flare.  This hypothesis is impossible to prove, but for
each case where we suspect it is present, we carry out a plausibility
analysis based on the energy budget. The energy of the radio-emitting
particles can be estimated from the radio flux, with various
assumptions, and this can be compared to the X-ray energy actually
measured. If the Neupert effect hypothesis is accepted, this also
places a constraint on the possible combination of magnetic field strength and
electron power-law index.  We outline the argument in some detail
below (Sect.~\ref{energybudget}).

If radio events occur which have no X-ray counterpart, the most likely
explanation is that the observed microwave flux arises from some 
process with low intrinsic energy in the electrons, most probably
an electron-cyclotron maser.  Such emission is
expected to be highly circularly polarized.
From the observing frequency, the local
magnetic field strength can be estimated.

Cases in which an X-ray flare is observed to have no discernable radio
counterpart are more problematic to explain. Some form of heating
process must be occurring, but any emission from accelerated particles
is somehow hidden from our view. Radio synchrotron emission from very
high energy particles is highly anisotropic and is predominantly
directed perpendicular to the field, $\eta_{\nu} \propto
\sin{\theta}^{(\delta+1)/2}$, where  $\eta_{\nu}$ is the
  emissivity, $\delta$ is the power-law slope of the electron
distribution, and $\theta$ is the angle between the line of sight
  and the local magnetic field (Dulk 1985).  In the case of low
energy electrons, the emission is directed parallel to the field, with
$\eta_{\nu} \propto \cos{\theta}^{2}$ for the fundamental, and in the
intermediate case of gyrosynchrotron emission from mildly relativistic
particles, the emission is broadly peaked perpendicular to the field
($\eta_{\nu} \propto \sin{\theta}^{-0.43 + 0.65\delta}$). The apparent
absence of radio emission accompanying some X-ray flares may point to
the heating particles being unusually low or unusually high in energy,
with an appropriate field geometry.

\subsection{Kinetic energy budget for radio flares}
\label{energybudget}

We follow here the analysis laid out in detail in G\"udel et al (2002a)
for a flare event of $\sigma$ Geminorum.  In outline, we first
estimate the number of accelerated particles produced and the total
energy contained in the distribution, based on the radio flux
observed, then compare this to the energy radiated in X-rays. A number
of assumptions must be adopted in making the estimate of the available 
kinetic energy. 

Non-thermal electrons in solar and stellar flares typically have a power law
energy distribution,
\begin{equation}
n(E,t) = \frac{N(t)\left( \delta -1 \right)}{E_0}(E/E_0)^{-\delta}, 
\end{equation}
 where $N(t)$ is the instantaneous number density of electrons of all
energies and $E_0$ is a cutoff energy for the distribution of
non-thermal electrons, which we chose to set at $10~keV$, a value
probably lower than typical solar flares which have $E_0 \sim 10-20
keV$ (Dennis 1988). We assume, based on solar analogy that the power
law index lies in the range $2.0 < \delta \le 3.5$ and $E_0 = 10$~keV
(see Dennis 1988). The total instantaneous kinetic energy  of the
  distribution is then
\begin{equation}
E_{kin}=N(t)V(t)\frac{\delta-1}{\delta-2}E_0 ,
\label{energyeqn}
\end{equation}
where $N(t)$ is the number density of electrons with energy above $E_0$,
and $V(t)$ is the source volume. If the source were optically thin, the observed radio flux over all 
bands would be, 
\begin{equation}
f_R(t) = \frac{\eta V(t)}{d^2} ,
\label{radiofluxeqn}
\end{equation}
where $\eta$ is the emissivity. In practive, we observe the radio flux
in a narrow band at 5~GHz (or 8.3~GHz in the case of
UV~Cet). Furthermore, the emission will probably not be entirely
optically thin. We correct for this by using some assumptions about
the usual nature of stellar coronal spectra. In many flares, the
turnover from optically thick low frequencies to optically thin high
frequencies occurs around 10-20~GHz, although can be as low as a few
GHz (Morris et al. 1990). The spectral index is typically $+1$ on the
optically thick low frequency side and between $-0.5$ and $-1$ on the
optically thin high frequency side. We can therefore estimate that at
5~GHz the intrinsic radio emission is in the region of 1~-~10 times
that observed. We assume a factor of 5 in the estimates which follow,
which corresponds to assuming the peak frequency $\nu_{peak} \sim 10$,
and spectral indices of +1 and -1 for the optically thick and
optically thin sides respectively.

The gyrosynchrotron emissivity for the {\em X}-mode is given by   
\begin{equation}
\begin{array}{lll}
\eta_x & =      & 3.3\times10^{-24} 10^{-0.52\delta} B N(t) (\sin{\theta})^{-0.43+0.65 \delta}  \\
       & \times & \left(\frac{\nu}{\nu_B}\right)^{1.22-0.9 \delta} , \\
\label{emisseqn}
\end{array}
\end{equation}
where $B$ is the magnetic field and $\nu_B$ is the electron
gyrofrequency (Dulk \& Marsh 1982). In what follows, we 
  routinely  consider values of $B$ between 20 and 200~G and set
$\theta$, the viewing angle, to 60$^{\circ}$, since most of the
emission arises from electrons with large pitch angles. We further
assume the {\em O}-mode emissivity is roughly equal to the {\em
  X}-mode emissivity, so that $\eta \approx 2 \eta_x$.

As a flare evolves, the number of accelerated particles is incremented
by ongoing acceleration processes, whilst the existing population
decays with some timescale $\tau$.  Thus
\begin{equation}
\frac{d[N(t)V(t)]}{dt} = \frac{d[N_+(t)V(t)]}{dt} - \frac{N(t)V(t)}{\tau},
\end{equation}
where $N_+$ is the total number of particles added by acceleration.
This equation may be integrated  in time twice 
over the duration of the flare to obtain the total number of injected
electrons.  If, after the flare, the number of additional
  particles returns to preflare levels, (which can be judged to
  be the case if the flux returns to preflare levels) the first
term on the  right hand side will not contribute to the
  integral and the total number of injected electrons is
\begin{equation}
N_{tot}=\frac{1}{\tau} \int^T_{T_0} N(t)V(t) dt .
\label{totplseqn}
\end{equation}
Combining equations~\ref{radiofluxeqn}, \ref{emisseqn} and~\ref{totplseqn}
we obtain
\begin{equation}
N_{tot}=1.55\times10^{19}10^{3.49 \delta}B^{0.22-0.9\delta}\frac{d^2}{\tau}\int_{T_0}^{T} f_R(t)dt .
\end{equation}
An appropriate value for the timescale $\tau$ is estimated from the
observed radio variations during the event. In practice, we have
usually adopted half the decay time for the radio flare, unless
shorter timescale variations are evident during the decay phase.
Equation~\ref{energyeqn} can then be used to determine the available
energy.

\section{Results and discussion}

\subsection{AD Leo}
\label{adleosection}

\begin{figure*}[!t]
  \begin{center}
   \vbox{
    \psfig{file=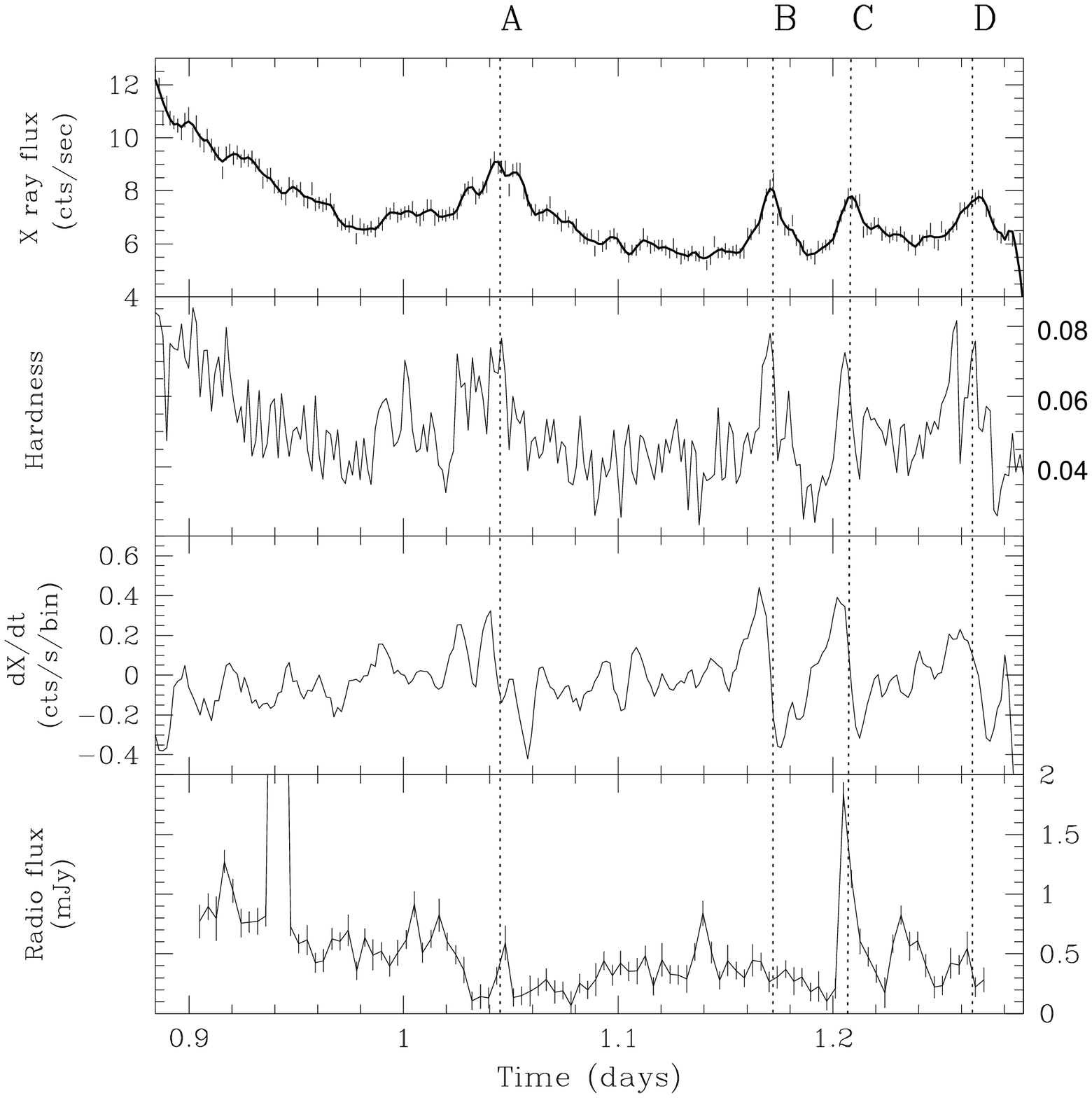, width=9.cm, height=9.cm}
\caption{Lightcurves for AD~Leo. From top: EPIC lightcurve, flare
hardness ratio, time derivative of the X-ray lightcurve, and radio
lightcurve at 5~GHz.  The time axis is labelled in days after IAT midnight
on 14.05.2001 (IAT differs from UT by the addition of 32 seconds). The identified X-ray flares are 
marked with dotted lines and labelled at the top of the plot.}
\label{adleoall}
\vspace{0.5cm} 
\hbox{
    \psfig{file=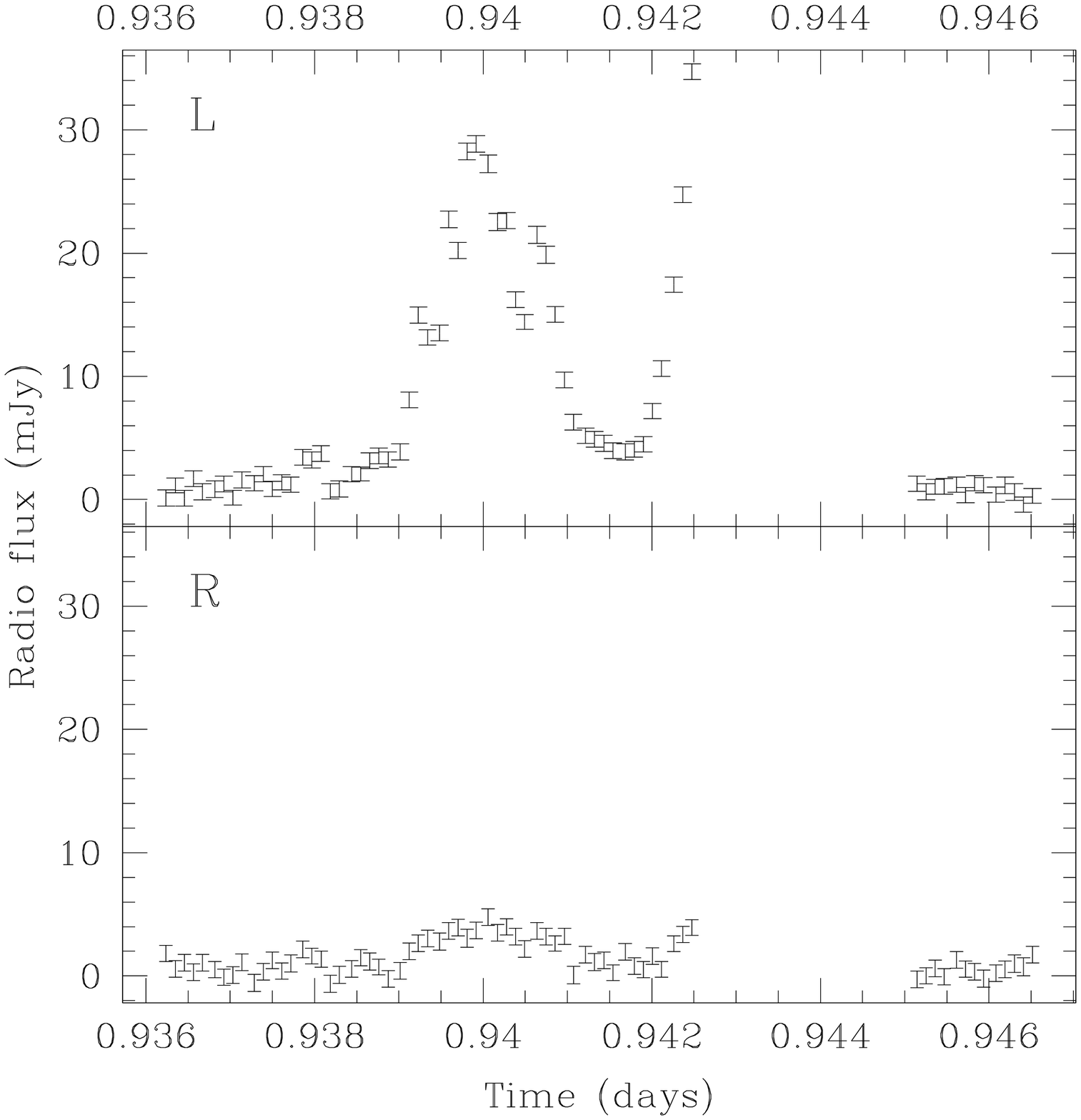, width=9.0cm, height=9.0cm} \hspace{0.3cm}
    \psfig{file=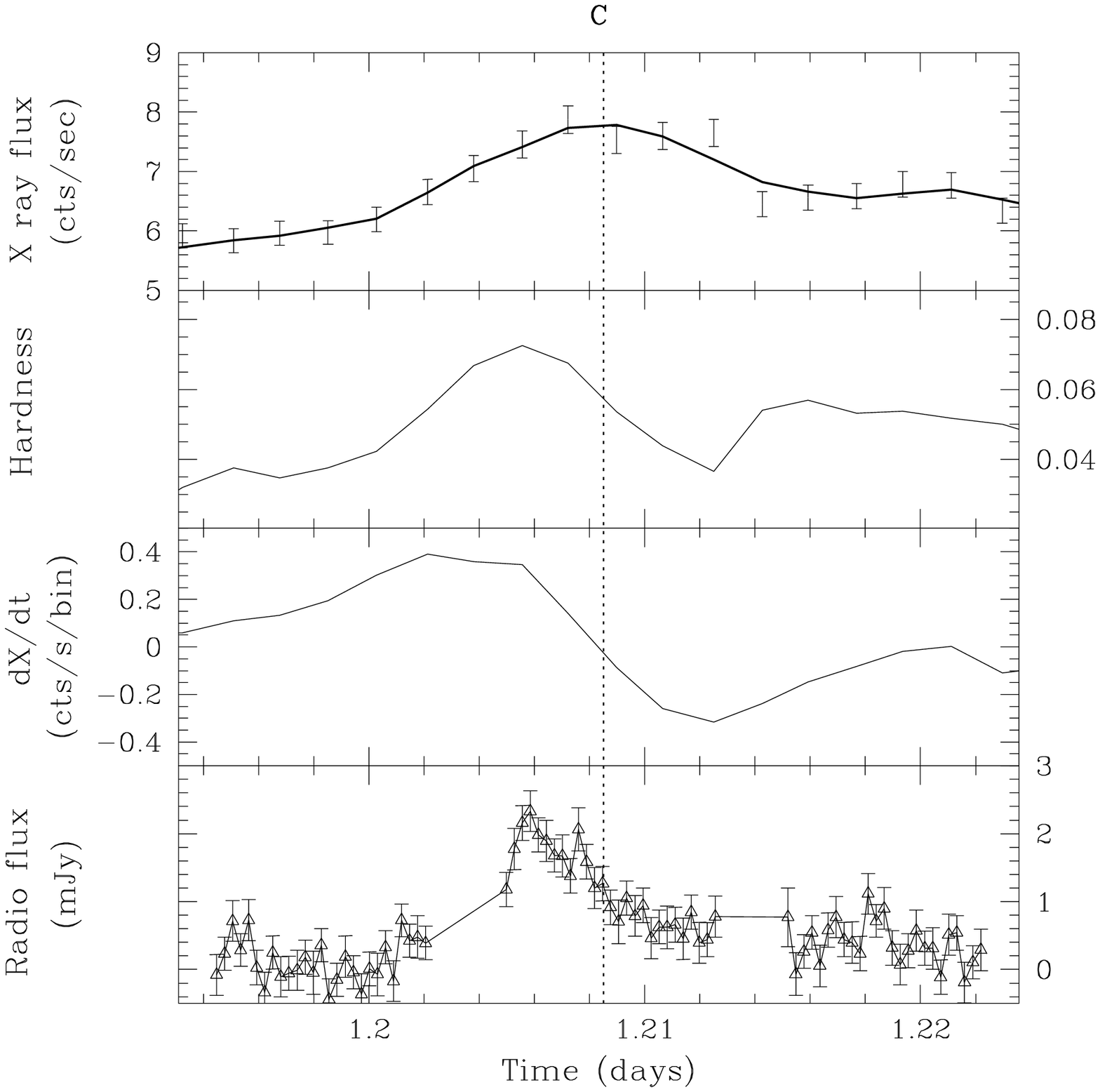, width=9.0cm, height=9.0cm}
    }}
  \end{center}
\caption{Left: Closeup of the radio lightcurve for the first radio
event.  The top panel shows the L flux, and the lower panel the R
flux. The gap is the break between scans, during which time the phase
calibrator was observed.  Right: Closeup of the event at 1.2 days
(event~C), showing the relative timing of the X-ray flux, derivative
of X-ray flux and radio flux.}
\label{adleocloseups}
\end{figure*}

The {\em XMM-Newton}~EPIC~pn and VLA 5~GHz lightcurves are shown in
Figure~\ref{adleoall}, together with the X-ray hardness ratio and the
first derivative of the X-ray lightcurve.  The top panel shows the
{\em XMM-Newton}~pn lightcurve, and a smoothed version of this made using a
Savitzky-Golay smoothing algorithm. The second panel shows the
X-ray hardness ratio for the flare. 
The third panel shows an estimate of the time
derivative of the X-ray lightcurve, also obtained using the
Savitzky-Golay filter. The bottom panel shows the radio lightcurve at 6cm.

Four distinct bright X-ray flares can be observed in the EPIC
lightcurve (Fig~\ref{adleoall}), and have been labelled A,B,C and D
for ease of reference.  Additionally, the beginning of the lightcurve
shows the decay phase of an earlier large outburst.  Apparently
continuous low-level radio activity is present from the beginning of
the VLA observation up to approximately 1.02~days. A large, shortlived
radio flare occurs at around 0.94~days, and has no clear counterpart
in the X-ray lightcurve. This flare is truncated on the scale in the
lower panel of Figure~\ref{adleoall}.  In the left hand panel of
Figure~\ref{adleocloseups} we show a higher time resolution plot of
the radio lightcurve at the time of this early radio flare. The
left-hand circularly polarized (LCP) flux is shown in the top panel,
and the right-hand polarized (RCP) in the lower panel.  This radio
flare is almost 100\% left-hand circularly polarized, suggesting that
this is a coherent emission process requiring fewer accelerated
particles and apparently leading to negligible chromospheric
heating. The main candidates for the emission process are either an
electron cyclotron maser, arising from a population of relativistic
electrons trapped in a magnetic flux tube, or plasma maser
emission. The former hypothesis allows us to directly estimate the
magnetic field strength in the emitting region, provided we assume the
emission occurs at the fundamental gyrofrequency
\begin{equation}
\nu_c = \frac{eB}{2 \pi m_e c} \approx 2.8\times 10^6 B.
\end{equation}
For our observations (6~cm $\equiv$ 5~GHz), this implies 
$B\sim1.8$~kG. This would drop by a factor of two 
if the emission in fact occurred at the first overtone. 
Plasma maser emission 
would occur at the fundamental plasma frequency
\begin{equation}
\nu_p = \left( \frac{n_e e^2}{\pi m_e} \right)^{\frac{1}{2}} \approx 9000 n_e^{0.5},
\end{equation}
which then implies that the density of the emitting material would be
$\sim 3\times10^{11}$~cm$^{-3}$, which is higher than values typically
found for stellar coronae, but not dramatically so.  Applying the
  energy budget argument laid out in Sect.~\ref{overview}, for the
  value of the magnetic field suggested above, the available kinetic
  energy of the accelerated particles is found to be of order
  $10^{29}$~erg, one or two orders of magnitude less than the radiated
  energy of most observed X-ray flares (see below).

A time-symmetric X-ray brightening occurs at about 1.04~hours
(event~A).  The early part of this feature occurs whilst the radio
flux is enhanced, apparently by a succession of modest flares, in the
early part of the lightcurve (up to about UT=1.03~days). A small radio
flare occurs at around the time of the peak of event~A.

\begin{figure}[!h]
  \begin{center}
\vbox{
     \psfig{file=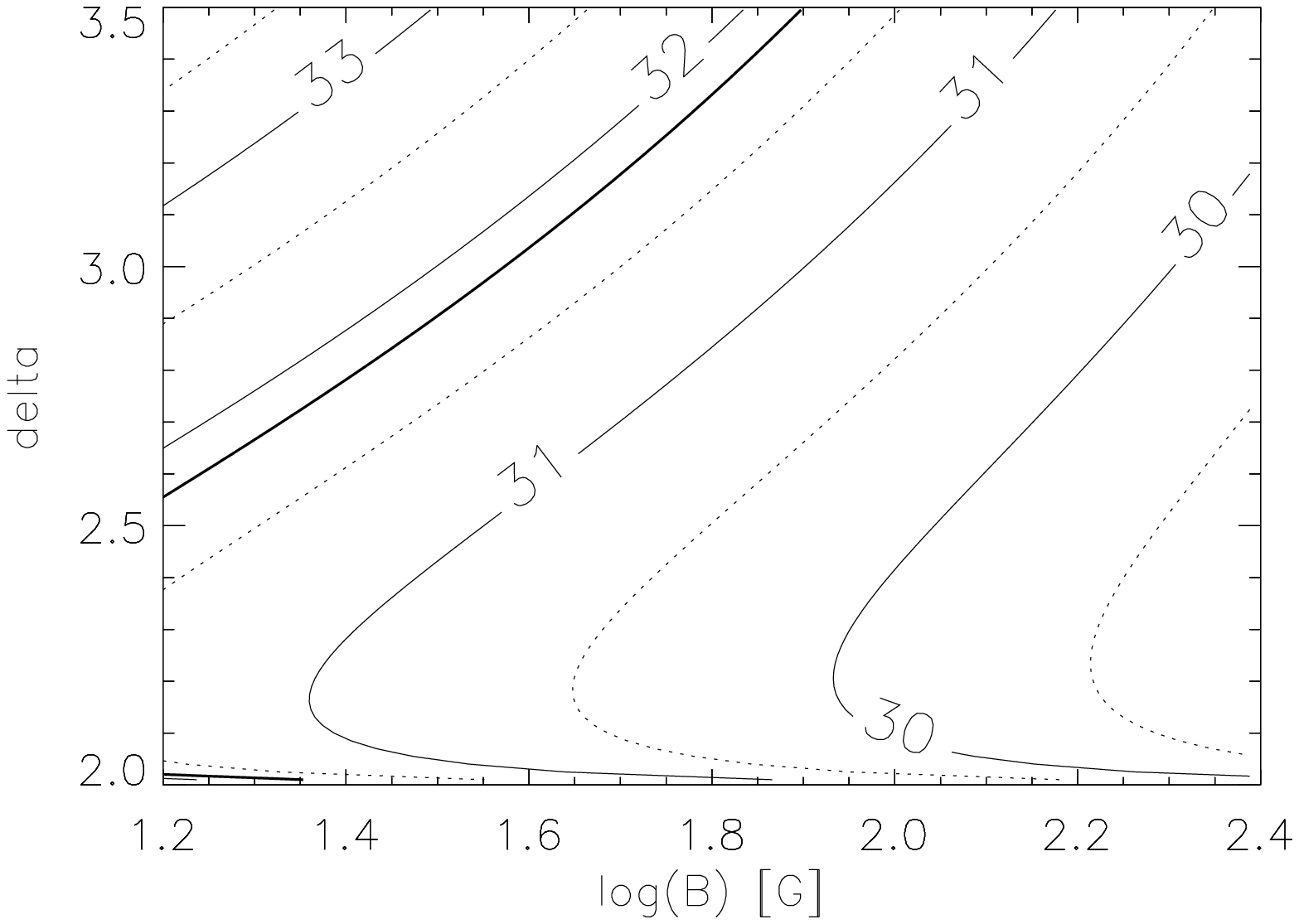, width=8cm, height=6cm}
     \psfig{file=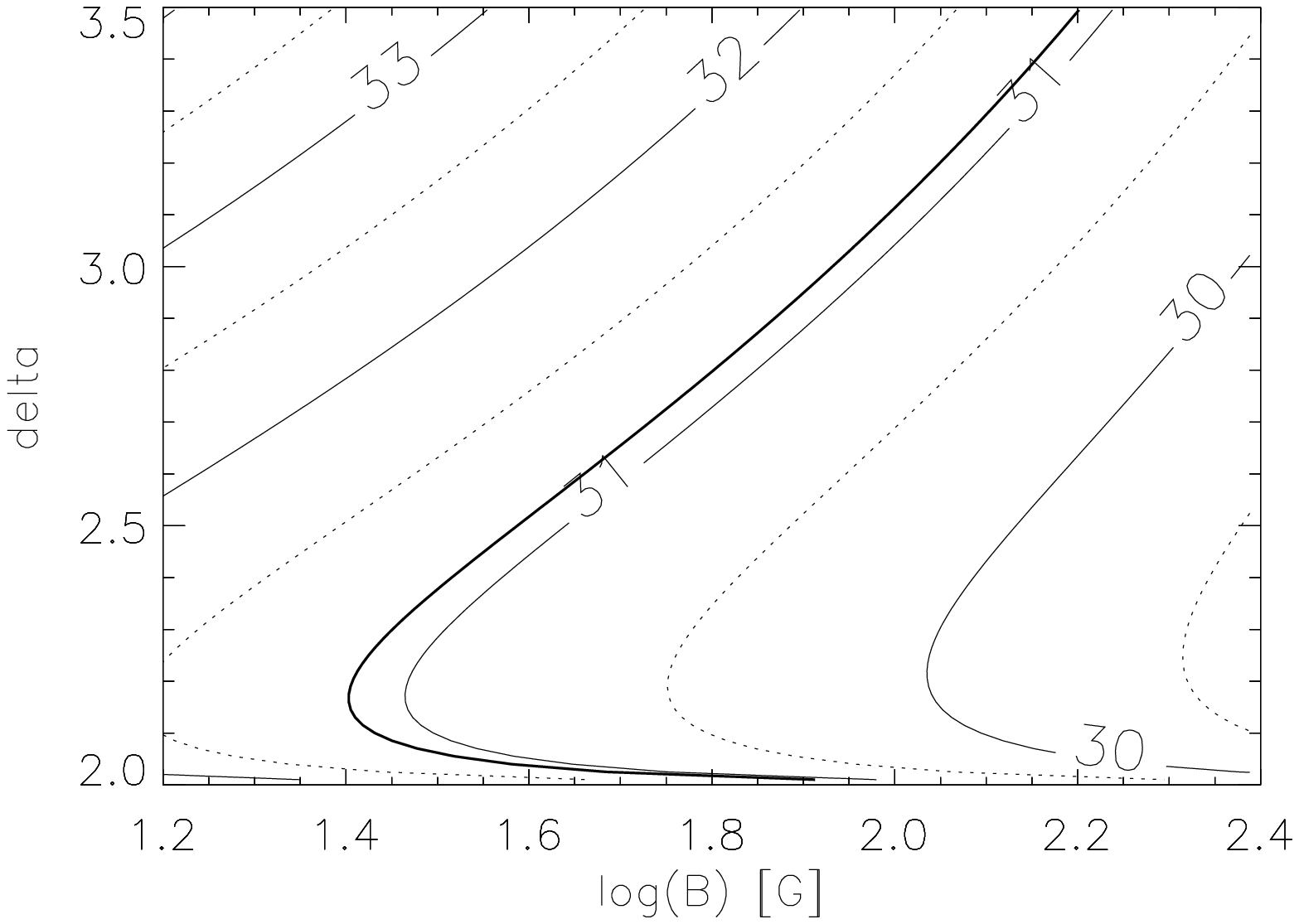, width=8cm, height=6cm}
}
  \end{center}
\caption{Top: Estimated kinetic energy of the emitting particles in the AD
Leo flare at T=1.05~days (event A) as a function of $\delta$ and
$B$. The solid contours show values of log(energy). Half decades are
shown with dotted lines. The contour corresponding to the estimated
radiated X-ray energy (approximately $7\times10^{31}$~erg) is marked with a
bold solid line.
Bottom: The equivalent plot for event C. }
\label{adcontours}
\end{figure}

To estimate the total energy emitted in the X-ray event, we have used
a standard Raymond-Smith model for a thermal plasma, with
parameters typical of M dwarfs (kT=1~keV, $n_{abs} \sim
10^{18}$~cm$^{-2}$), to estimate the mean energy per photon.  We
estimate from the observed count rate that around $7\times10^{31}$~erg
are radiated in the X-ray flare. The available energy from the
accelerated particles estimated from the radio flux as described above
in Section~\ref{energybudget} is plotted as a function of $B$ and
$\delta$ in Figure~\ref{adcontours} (upper panel).  The radiated X-ray energy accounts
for a high fraction of the available energy, even with 
helpful assumed values of $B$ and $\delta$.  Also, although the small flare 
at T=1.05~days is simultaneous with the peak of the X-ray flare, 
the brightening in X-rays is sustained over a much longer timescale, 
so that the radio activity does not correlate 
very closely with the overall X-ray behaviour. It therefore seems
likely that the total radio flux associated with the heating particles
is not visible to us. One way this might occur is if the radio
emission is strongly anisotropic, as discussed above in
Section~\ref{overview}.  Strongly anisotropic emission is expected for
synchrotron emission from very high energy particles, and the flux
would be mostly directed perpendicular to the field lines. The lack of
a radio counterpart to event~A would then imply that the field loops
lie approximately along the line of sight for most of their
length. 

A small radio flare occurs at around T=1.14~days.  This radio flare
was found not to be significantly polarized.  Nevertheless, there is
no X-ray counterpart.

Event B, at UT=1.17~days, has no clear radio counterpart in the
overall lightcurve. A closer examination of the radio lightcurve at higher time
resolution revealed no significant short timescale brightening at this
time, although there were low-level fluctuations in the lightcurve
greater than the noise level. None of the small peaks exceeded 1~mJy
in brightness. The radiated X-ray energy is approximately
$2\times10^{31}$~erg.  It seems highly unlikely that this energy is
provided by particles whose radio emission we observe. Anisotropic
emission seems a more likely explanation.

The third radio flare is approximately simultaneous with the third
X-ray flare, event C (Figure~\ref{adleocloseups},~right hand
panel). The close-up shows that the X-ray lightcurve in fact peaks
some 4 minutes after the radio peak. The gradient of the X-ray
lightcurve correlates well with the radio emission. As noted above,
this correlation is expected from the Neupert effect. We again follow
the steps outlined in Sect.~\ref{energybudget} to assess the
plausibility of the Neupert effect hypothesis.  We used values of
$\delta$ between 2.0 and 3.5 and $B$ between 20 and 200~Gauss. The
decay timescale is estimated from the radio lightcurve to be around
180~s. A contour plot of energy as a function of these parameters is
shown in Figure~\ref{adcontours}.  The total injected energy is
between $10^{29}$~erg and $10^{33}$~erg. Just over $10^{31}$~erg is
emitted in the X-ray flare. The Neupert effect hypothesis is therefore
broadly consistent with the energy budget in this case.
However, the radio flare was found to be strongly left-hand circularly
polarized ($\sim$~75\%). This implies that a large fraction of the
radio emission arises from processes other than gyrosynchrotron. If we
assume that the observed emission is a blend of gyrosynchrotron
emission, with a 20\% polarization, and emission from some coherent
process, which is nearly 100\% polarized, this would mean that about
30\% of the total emission would be gyrosynchrotron. There would still
be sufficient energy in this case to allow for the observed X-ray
flux.

Radio activity occurring just after event~C, at
about T=1.24~days, had no clear X-ray counterpart and was also
left-hand circularly polarized at the 50-100\% level. This is therefore 
most likely emission due to some coherent process.

A final X-ray variation occurs near the end of the observation (event
D). The total energy radiated in X-rays is approximately
$1.4\times10^{31}$~erg. The radio lightcurve is not particularly
well-correlated but does show some brightening at this time.

\subsection{AU Mic}

\begin{figure}[!h]
  \begin{center}
\vbox{
    \psfig{file=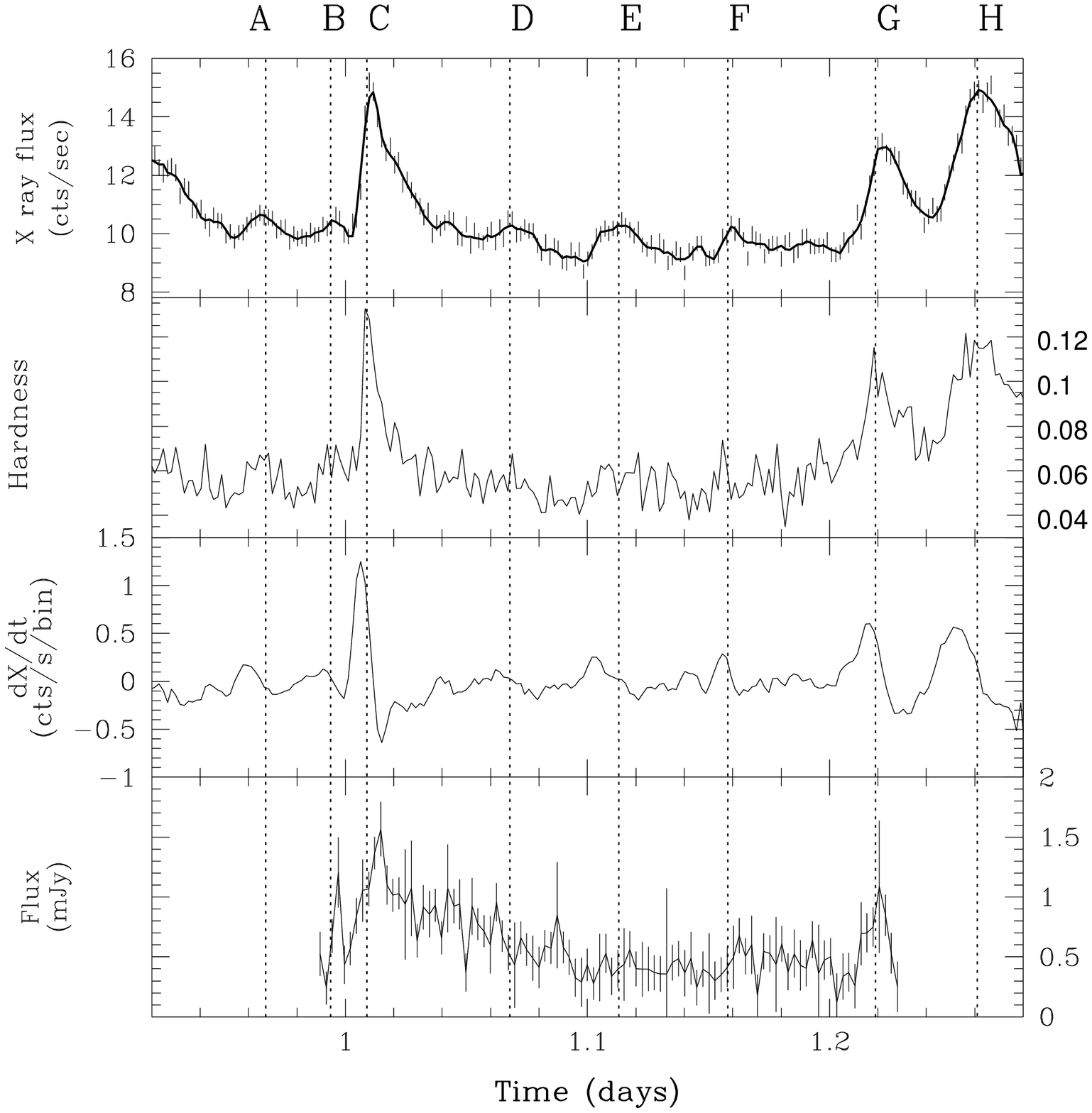, width=9.cm}
\psfig{file=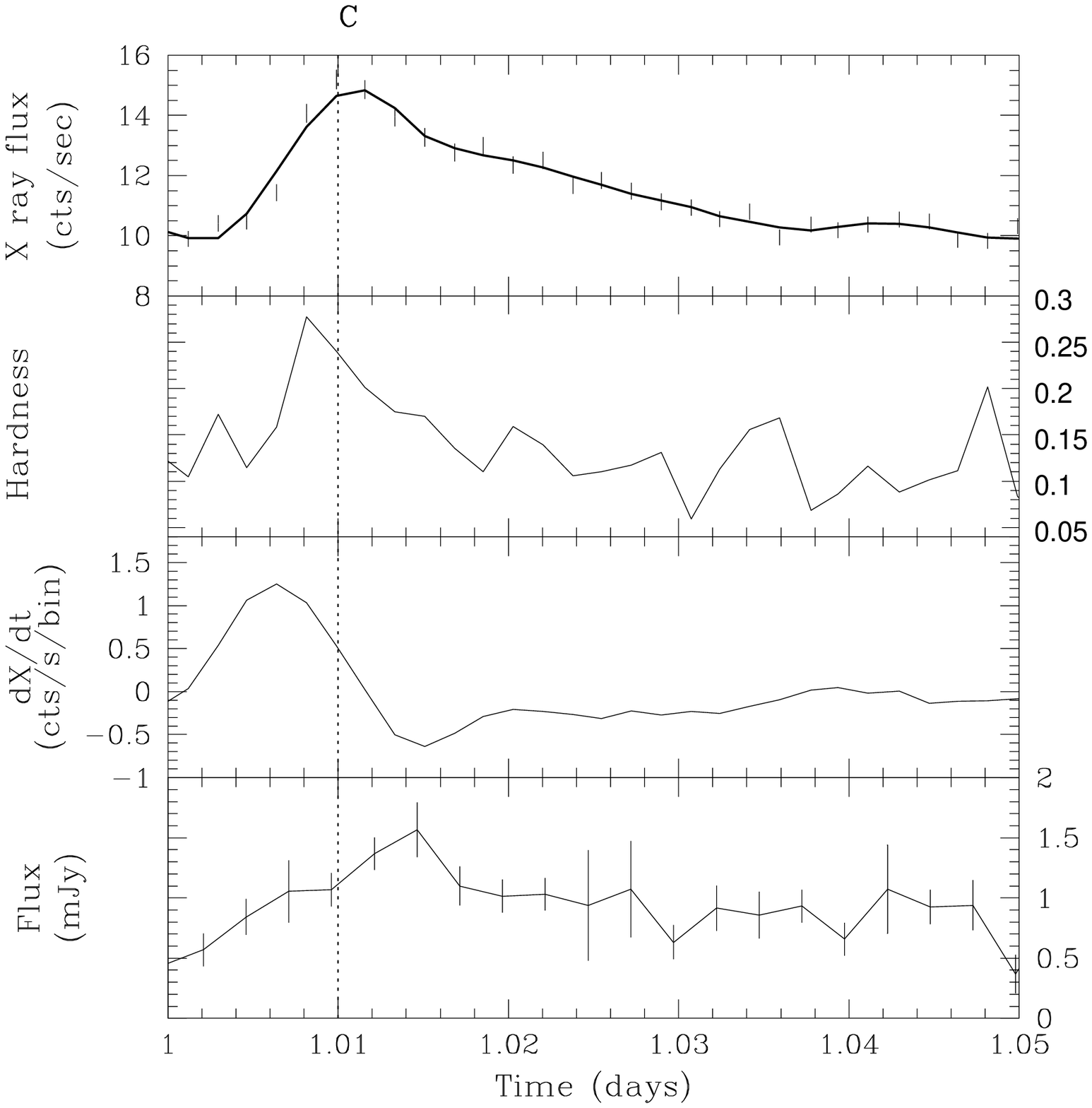,width=9.cm}
}
  \end{center}
\caption{Upper figure, from top to bottom: X-ray flux, hardness ratio,
derivative of X-ray flux and radio flux for AU Mic. 
The time axis is labelled in days since IAT midnight 
on 13.10.2000. Lower figure:
close up of lightcurves around event C.  }
\label{aumic}
\end{figure}

 \begin{figure*}[!ht]
 \hbox{
     \psfig{file=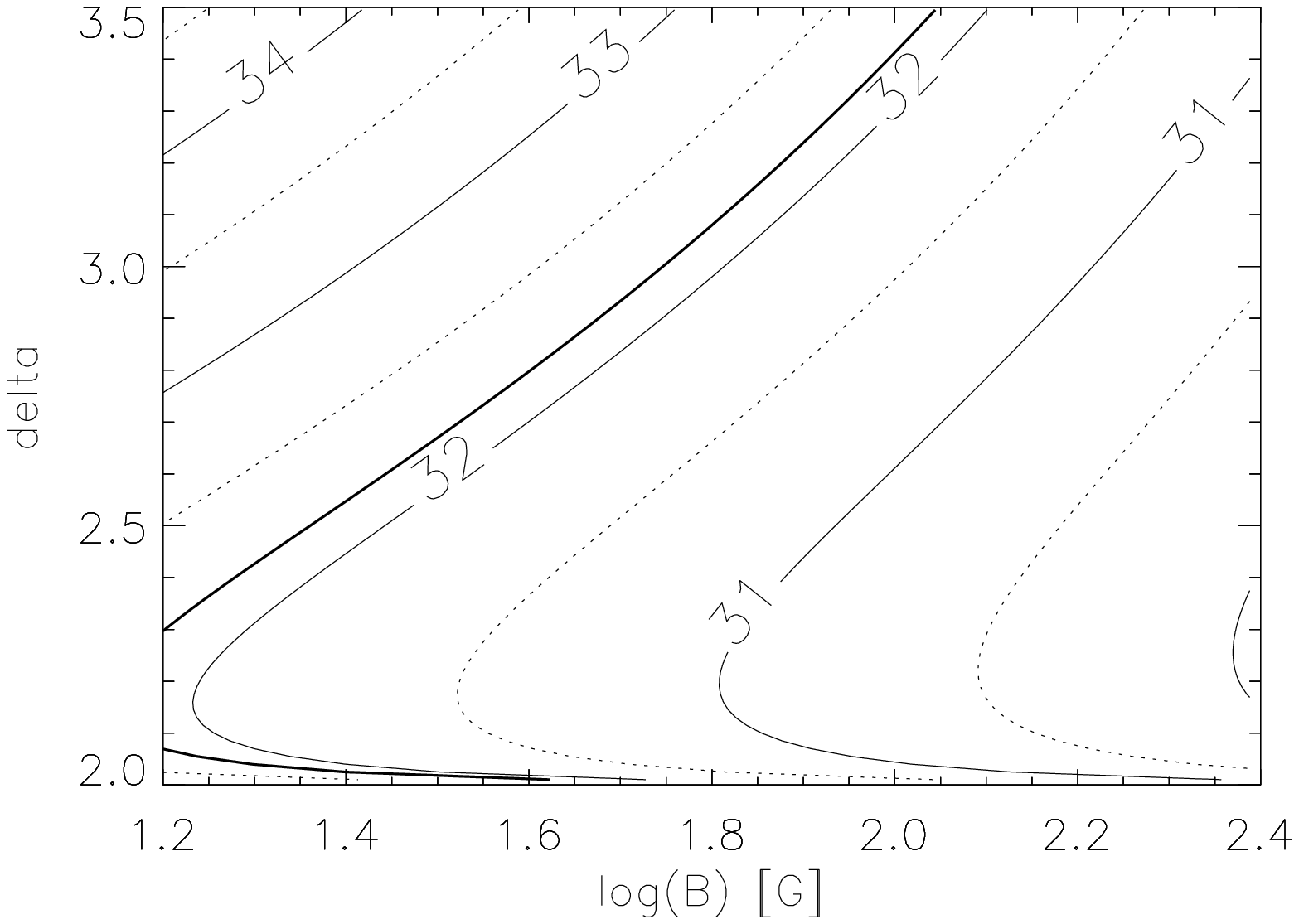, width=8cm, height=6cm}
     \psfig{file=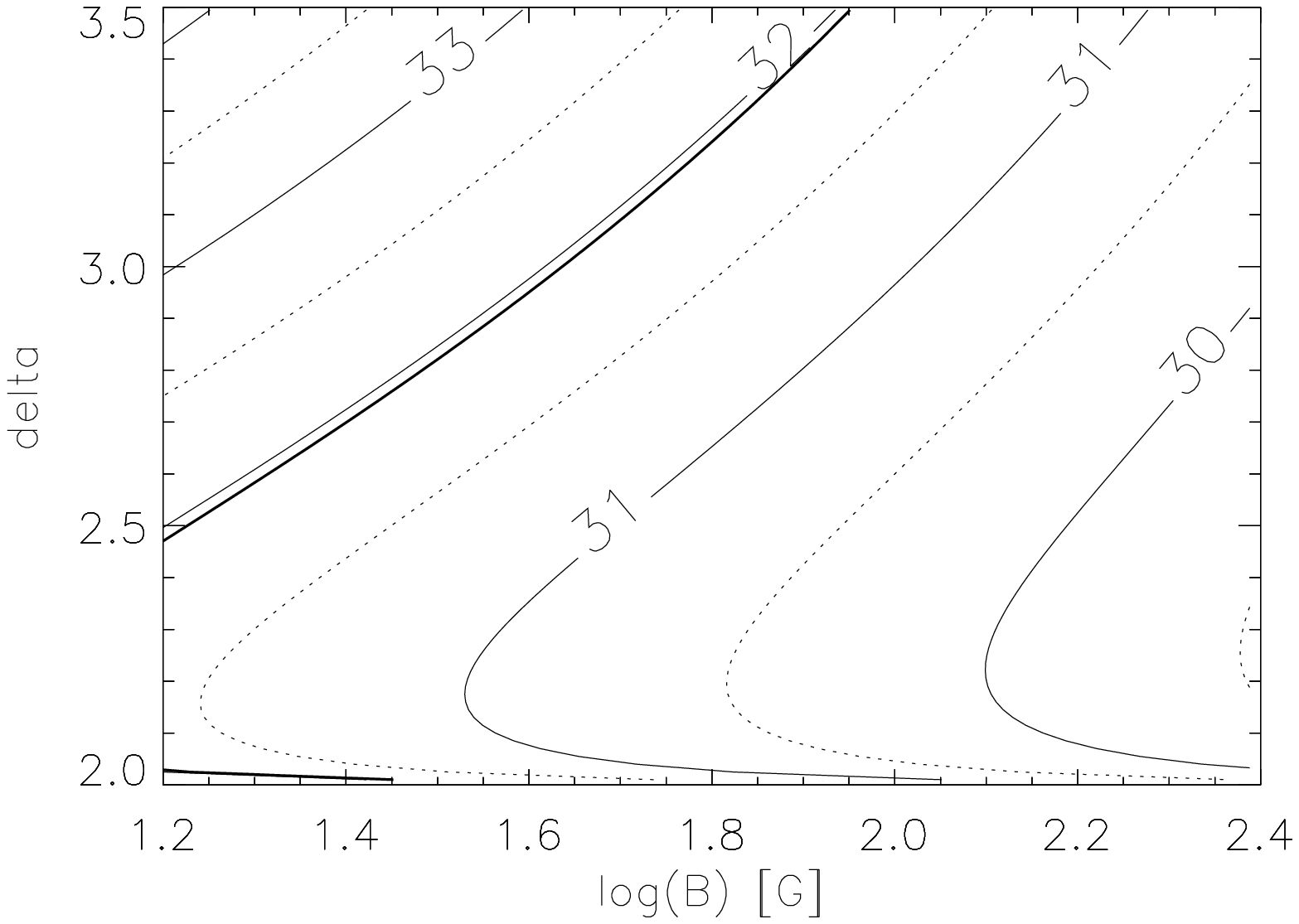, width=8cm, height=6cm}
 }
 \caption{Estimated kinetic energy of the emitting particles in the AU
 Mic radio flares corresponding to X-ray flares C (on the left) and G
 (on the right) as functions of $\delta$ and $B$. The solid contours
 show values of log(energy). Half decades are shown with dotted
 lines. The bold solid lines show the radiated X-ray energy. This was
 approximately 1.5$\times10^{32}$~erg for event C and
 9$\times10^{31}$~erg for event G.}
 \label{aucontours}
 \end{figure*}

The {\em XMM-Newton} pn and VLA 5GHz lightcurves are shown in Figure~\ref{aumic},
together with the hardness ratio and X-ray lightcurve derivative. We
have labelled eight distinct events in the X-ray lightcurve with
letters A-H. Event A occurs before the radio lightcurve begins, and
event B occurs only slightly after the onset of the radio lightcurve.
Nevertheless a sharp spike in the radio lightcurve occurs at about this time.

Event C seems to be correlated with the large radio flare at the
beginning of the radio lightcurve. This section of the lightcurve is
shown in greater detail in the lower panel of Figure~\ref{aumic}.
This first radio flare is assumed to last from the beginning of the
observation until $t=1.03$~days, when the radio emission dips
temporarily, although it could be argued that the decay phase in the
radio lasts much longer, until around UT=1.15~d.  Following the same
argument used for AD~Leo above, with an estimated decay timescale
$\tau$, of 500~s corresponding to the most rapid timescale seen in the
flare, at around 1.025 days, we estimate the energy available at the
flare site to be between $10^{30}$ and $10^{34}$~erg (see
Figure~\ref{aucontours}, left hand panel). The radiated X-ray energy is
estimated to be $1.5\times10^{32}$~erg (see Fig.~6 left-hand panel).  

A further small X-ray event, labelled D, occurs late in the decay
phase of the radio flare and has no clear radio counterpart.  Two
further small events in the X-ray lightcurve, E and F, have no clear
radio signature, although F may correspond to a slow increase in radio
flux at about this time. A larger flare at the end of the radio data,
event G, has a clear radio counterpart. Here, the total available
energy is approximately $10^{30}$--$10^{33}$~erg, while the X-ray
event radiates approximately $9\times10^{31}$~erg. The kinetic energy
of the accelerated particles is plotted as a function of $B$ and
$\delta$ in the right-hand panel of Figure~\ref{aucontours}.  The
X-ray energy needed is at the high end of the range of available
energies estimated from the radio lightcurve.  A larger X-ray flare,
labelled H, occurs after the end of the radio lightcurve.  None of the
radio variations showed a level of circular polarization above
approximately 20\%.

\subsection{AT Mic}

\begin{figure}
  \begin{center}
    \psfig{file=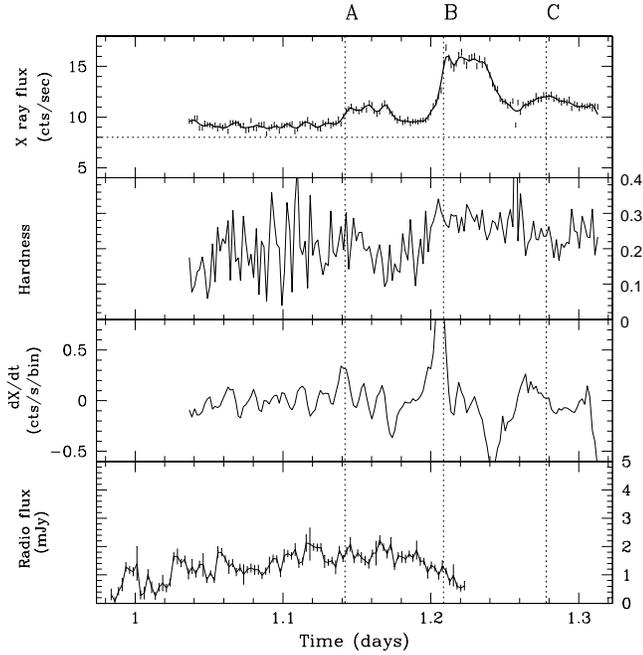, width=9.cm, height=9.cm}
  \end{center}
\caption{AT Mic lightcurves. From top to bottom, X-ray flux, hardness
ratio, derivative of X-ray flux, and radio flux. The time axis is
labelled in days since IAT midnight on 15.10.2000.  }
\label{atmic}
\end{figure}

Although the binary components of AT~Mic could be distinguished in
the VLA maps, it is difficult to produce separate lightcurves because
the VLA beam is highly distorted over short intervals. We have
therefore considered the total flux of both objects together.  The
lightcurves are shown in Figure~\ref{atmic}.  The radio flux increased
at the beginning of the observation, reaching a plateau of 
around 1.5~mJy, and then showed signs of a decrease 
towards the end of the observation. There was little or no rapid 
activity in the radio lightcurve once the plateau was reached. 
The flux was approximately 30\% LCP throughout. 

We have identified three events in the X-ray lightcurve (labelled A, B
and C).  A and B each show an unusual profile with a flat top and
approximately equal rise and decay phases, although event B is much
stronger than event A. Event C is a very gentle time symmetric
brightening which occurs after the radio observation had ended and will
not be discussed further.

The energy radiated in X-rays by event A is estimated to be
approximately $1.5\times10^{32}$~erg. The energy radiated by event B
is approximately $6\times10^{32}$~erg. No short-timescale radio
counterpart could be found for either event.

\subsection{UV Ceti}

\begin{figure}[h]
  \begin{center}
    \psfig{file=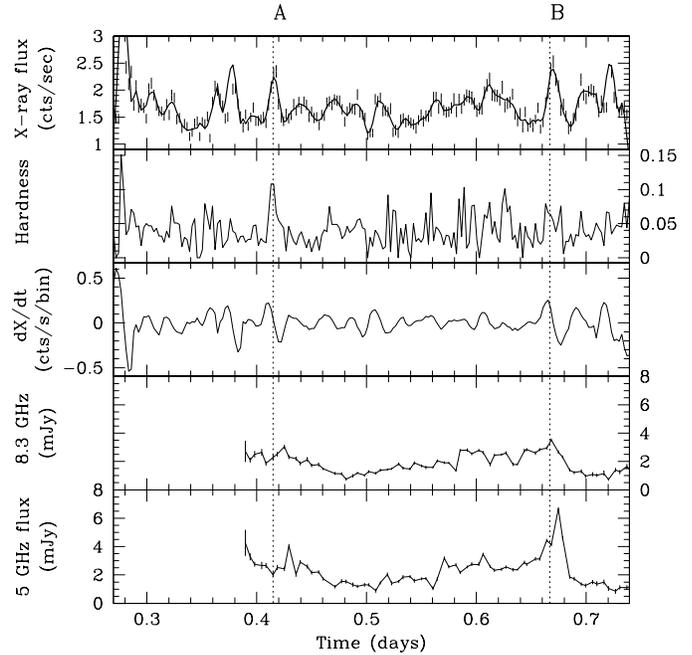, width=9cm}
  \end{center}
\caption{Lightcurves for UV~Ceti. From the top, the X-ray flux,
hardness ratio, derivative of the X-ray flux, 8.3~GHz flux and 5~GHz flux. 
The time is in IAT days after IAT=0h on 07.07.2001}
\label{uvceti}
\end{figure}

The components of UV~Cet were not distinguishable in the X-ray frames.
For this star, we obtained radio data at both 5 and 8.3 GHz.  These
are shown separately in Fig.~\ref{uvceti}. The X-ray lightcurve shows
almost continuous flaring activity. To avoid confusion, we have labelled
only two events in the figure, which are somewhat larger and more
distinct than the others and which correspond roughly in time to radio
activity. The radio lightcurves also show what may be a continuous
sequence of small flares. The lightcurves at the two different radio
frequencies are very similar, except for the large flare near the end
of the observations, which is seen strongly at 5~GHz and hardly
appears at all at 8.3~GHz.

\begin{figure}[h]
  \begin{center}
    \vbox{
    \psfig{file=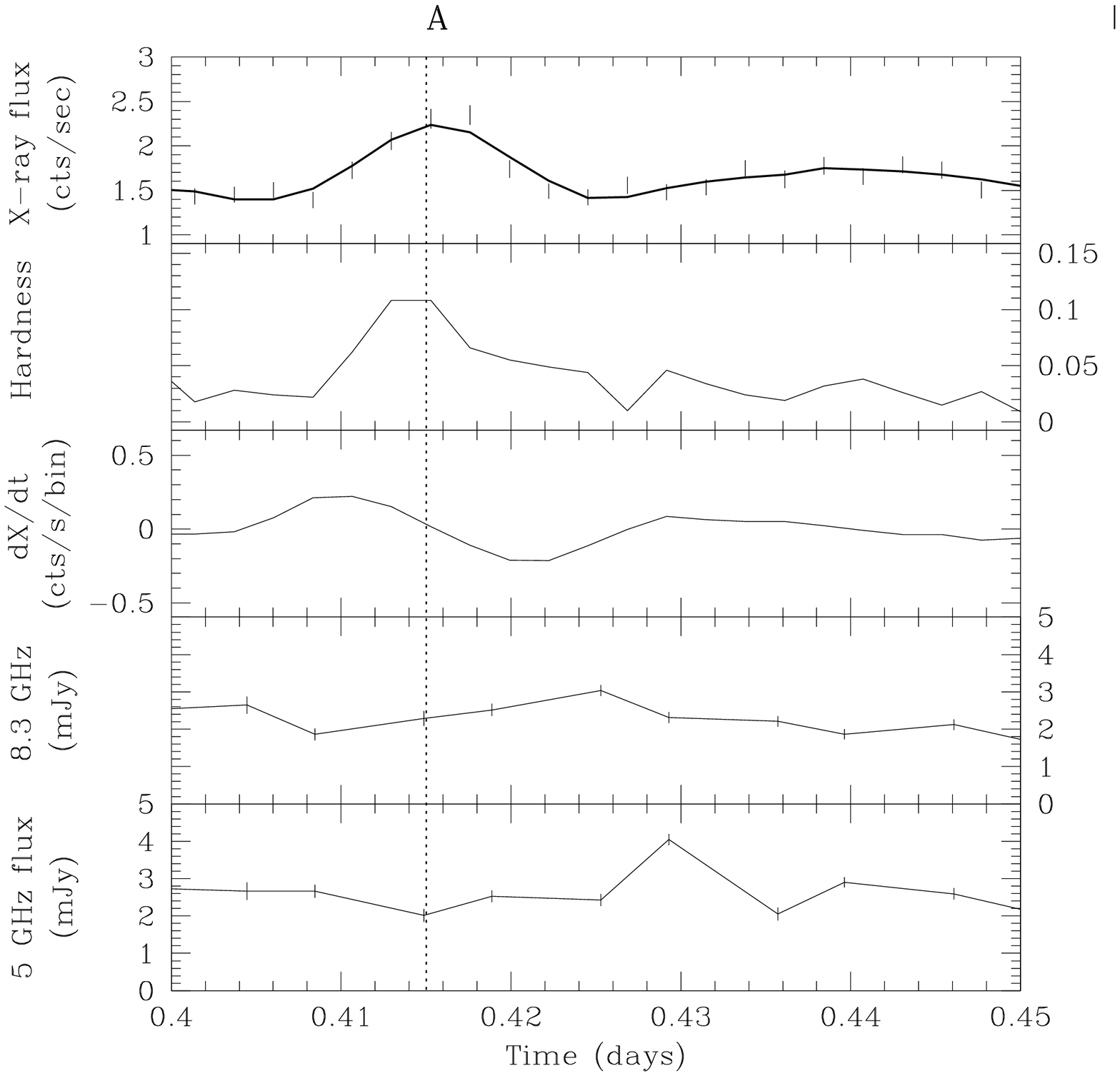, width=9cm}
    \psfig{file=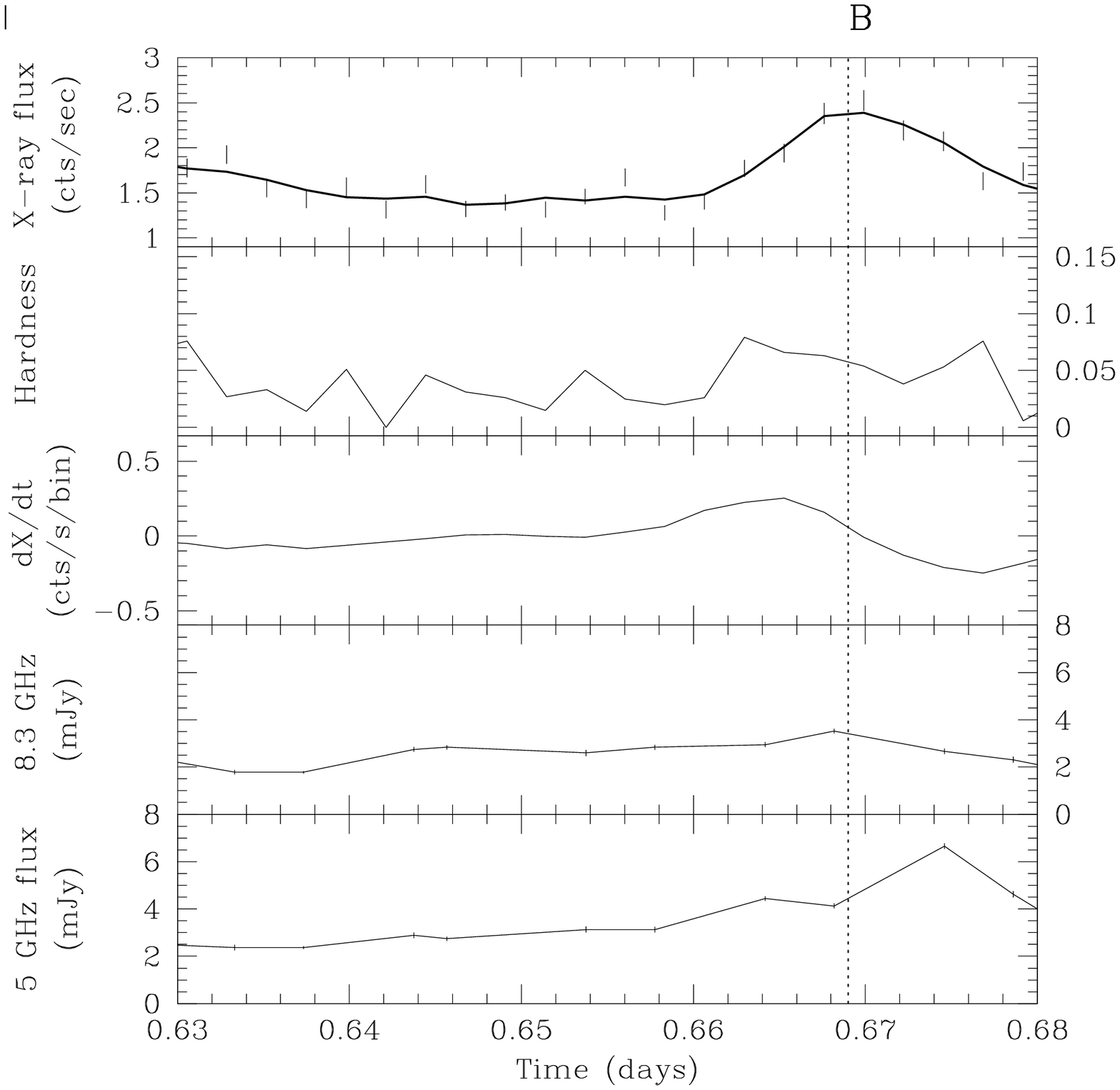, width=9cm}
    }
  \end{center}
\caption{Closeup of event A (top) and event B (bottom) for UV~Cet}
\label{uv_close}
\end{figure}

Event~A occurs just after the beginning of the radio observations.  A
small radio flare occurs shortly afterwards at 8.3~GHz, and a few
minutes later a small flare is seen at 5~GHz.  The X-ray flare
radiates just over $1 \times 10^{30}$ ergs.  This is in the middle of
the range of available energy as a function of $B$ and $\delta$ as
estimated from either radio band.  However, the correlation between
the radio flux and the derivative of the X-ray lightcurve is
highly questionable (Fig.~\ref{uv_close} upper panel).

Event B corresponds well to a local peak in the lightcurve at 8.3~GHz, and
precedes the larger flare seen at 5~GHz by several minutes (see
Fig.~\ref{uv_close}, lower panel). The radiated X-ray energy is around
$10^{30}$~erg, which lies approximately in the mid range of available
energies estimated from the 8.3~GHz radio flux and towards the low end
of energies available as estimated from the 5~GHz flux.

The average spectral index (defined by $F_{\nu} \propto \nu^{\alpha}$),
implied for the spectrum of UV~Ceti is
$\alpha=-0.26$. However, this becomes more steeply falling during the event B,
where most of the extra flux emerges at 5GHz. The spectral index at
the peak of the radio flare is $\alpha=-1.8$.  This type of sharply falling
spectrum is suggestive of gyrosynchrotron emission.

 \subsection{YZ CMi}

\begin{figure} 
\begin{center} 
\psfig{file=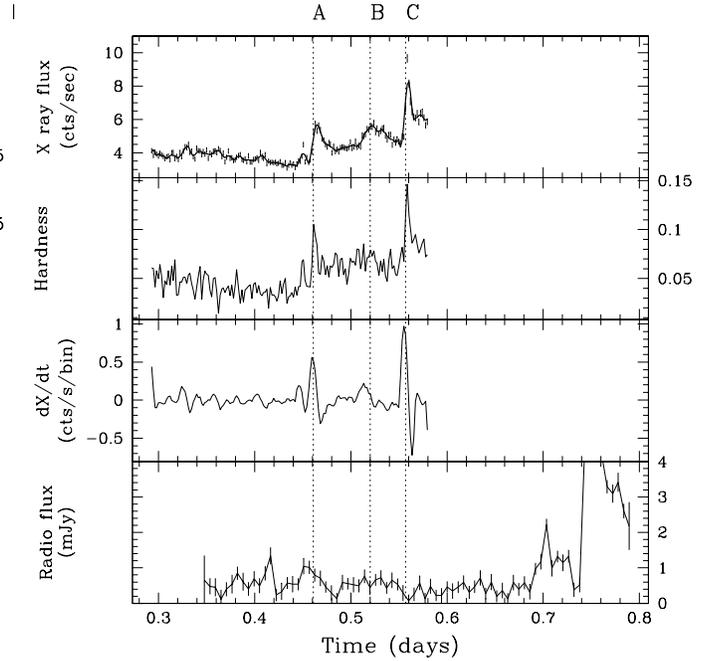, width=9cm}
\end{center} 
\caption{Lightcurves for YZ~CMi. The layout is the
    same as for previous cases for AD~Leo, AU~Mic and AT~Mic.
The time axis is labelled in days since 9.10.2000.}  
\label{yzcmi}
\end{figure}

The X-ray lightcurve of YZ CMi (Fig.~\ref{yzcmi}) shows a period of
quiescence followed by a series of flares towards the end of the
observation.  An increase in flux occurs at about 0.45~days, labelled
A, and is followed by a higher plateau, which leads slowly into
another increase just past 0.5 days (event B).  Finally, a flare
occurs just before the end of the observation at about 0.56~days
(event C). The radio lightcurve shows a small flare-like event,
at about 0.41 days, and possibly another small flare at 0.45 days
which may correspond to event A. 
Neither of these flares was found to show any significant circular
polarization. There is no clear counterpart to events B or C. 

The energy budget calculation was performed for the flare labelled
A. The X-ray flare radiates an estimated energy of
$3\times10^{31}$~erg. Event B appears very similar, with a more
gradual and symmetric X-ray event preceded by a small radio spike. The
energy budget calculation is very similar for this event as for event
A, with the X-ray energy being approximately
$3\times10^{31}$~erg. Event C appears not to have a corresponding
radio flare. A search of the high-resolution radio data also failed to
uncover a counterpart.

\subsection{Overall picture}

\begin{figure}
\psfig{file=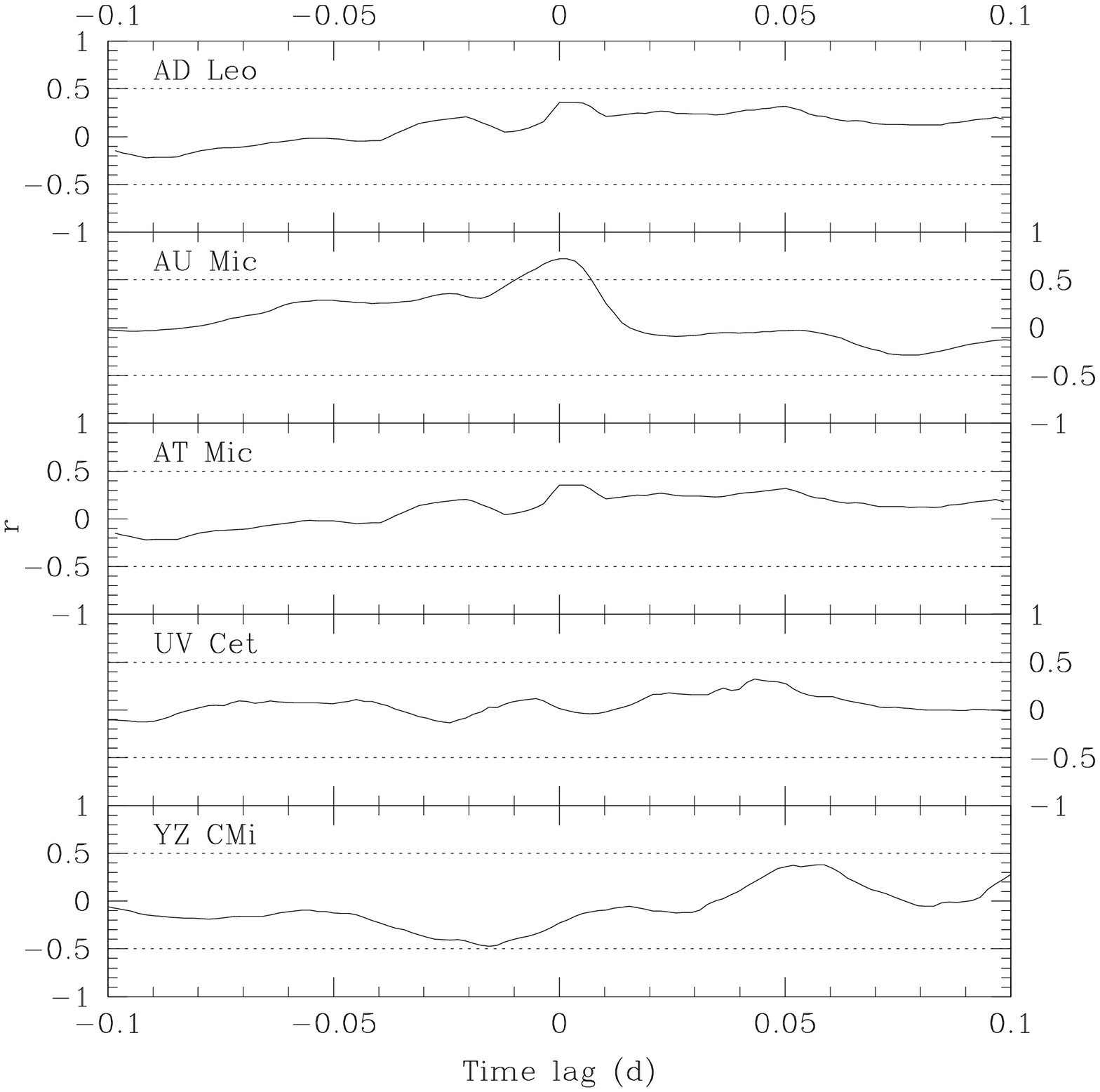, width=9.cm}
\caption{Cross correlation of the X-ray and radio lightcurves for the five 
targets. Correlations of $\pm$0.5 are indicated with dotted lines. 
The overall lightcurves do not correlate 
well, with the exception of AU Mic which has a peak correlation of 0.72.}
\label{overallcorrn}
\end{figure}

In Figure~\ref{overallcorrn} we plot the cross correlation of the
radio lightcurves with the X-ray lightcurves for each of the stars. 
The lightcurves do not show strong correlations over their entire 
duration, with the exception of AU Mic which has a correlation coefficient of 
0.72 at zero lag.

\begin{figure*}
\hbox{
\psfig{file=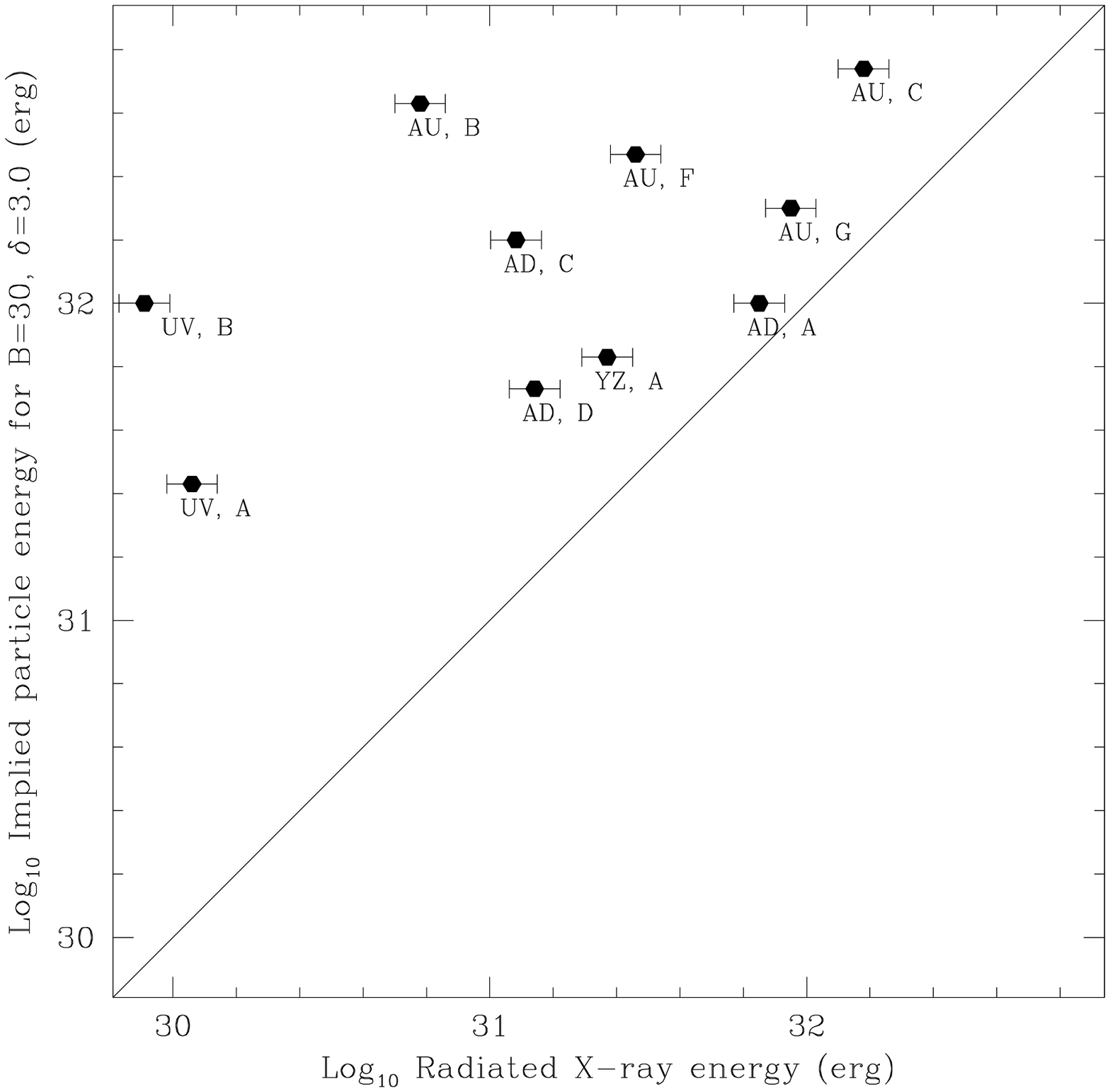, width=9.cm}
\psfig{file=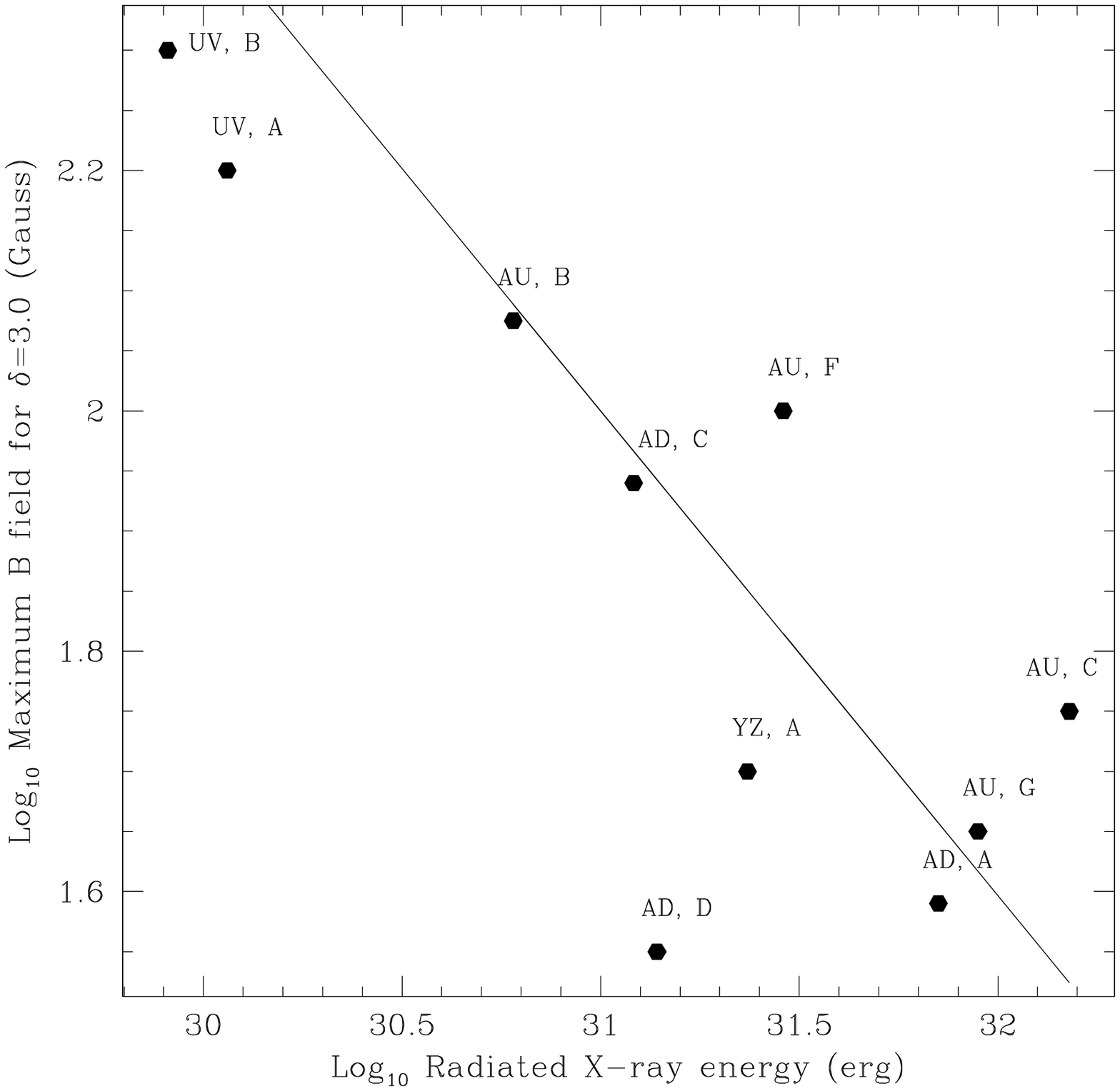, width=9.cm}
}
\caption{Left: Comparison of radiated X-ray energy with inferred
  particle energy for ten events in which plausible radio counterparts
  to X-ray flares could be identified.  Each point is labelled with
  the first two letters of the stars' name and the identification
  letter of the flare in question.  Right: The value of $B$ for each
  of the ten events for which the particle kinetic energy matches the
  radiated X-ray energy for $\delta=3.$. This is the maximum value of
  $B$ possible with that assumption. The straight line shows the
  relation $B \propto E_X^{-0.4}$, which is expected in the event that
  the radio flux is uncorrelated with the X-ray flux. The events AD A,
  AD C, AU C and AU G are shown in detail in the various contour plots
  above.}
\label{radxrcorrn}
\end{figure*}

In the left-hand panel of Figure~\ref{radxrcorrn} we show the particle
kinetic energy as estimated from the radio data plotted against the
radiated X-ray energy for a total of ten events for which a plausible
radio counterpart to an X-ray flare was observable. The particle
kinetic energy is a function of the local magnetic field and the
electron energy distribution power-law index, so it is necessary to
choose fiducial values of these quantities in order to calculate a
representative energy.  We have chosen $\log(B)=1.5$, which is
equivalent to $B=31.6$~G, and $\delta=3.0$, which results in total
particle kinetic energies larger in every case than the radiated X-ray
energy. The dependence of the estimated energy on the choice of values
for $B$ and $\delta$ can be estimated from the various contour plots.
To decrease the inferred particle energy by one order of magnitude
requires in each case an increase of approximately a factor of two in
$B$, or alternatively a decrease in $\delta$ to around 2.3.  A loose
correlation between the two energies plotted in Fig.~\ref{radxrcorrn}
is visible by eye. The Pearson correlation coefficient is $0.52$. A
non-parametric Kendall test reveals rank correlation with a
significance at the 1.5$\sigma$ level. The correlation is therefore
not significant.

In the right-hand panel of Fig.~\ref{radxrcorrn} we show for each
event the value of $B$ implied if the available kinetic energy is
equal to the radiated X-ray energy, for the case that $\delta=3$. This
is then the maximum value of $B$ which is possible with that value of
$\delta$, and must underestimate the true value since the heating and X-ray emission
process will not be 100\% efficient. The straight line shows the
relation $B \propto E_{X}^{-0.4}$, which would be expected
if the observed radio flux were independent of the observed X-ray
flux. No systematic deviation from this relation can be seen.

\section{Conclusions}

We have observed five flare stars simultaneously in X-rays with the
{\em XMM-Newton} EPIC cameras and with the VLA.  Comparisons of the
behaviour at these two regimes can reveal relationships between the
radio and X-ray emission which can often be interpreted in terms of
standard models developed for solar flares. In particular,
nearly-simultaneous radio and X-ray flares are suggestive of the
Neupert effect, in which the same population of accelerated electrons
is responsible for the synchrotron emission seen in radio, the heating
of material which then emits in hard X-rays, and the evaporation of
chromospheric material which then emits soft X-rays.

We identified a total of 17 large, distinct X-ray flares during the
time periods when radio observations were also available.  Of these,
ten had possible counterparts in the radio lightcurve, and of these
ten four showed good correlation between strong events in X-ray and
strong events in the radio lightcurve. An energy budget calculation
was performed for each of the ten cases, in which the total energy
available from gyrosynchrotron emitting electrons was estimated for
various values of magnetic field strength and power-law slope, and
compared to the total energy radiated in X-rays. The radiated X-ray
energy should be lower than the estimated available particle energy
for the Neupert effect to be a viable hypothesis to explain the 
apparently related flare events. In all these cases the Neupert effect
hypothesis was found to be credible on energy budget grounds. 
Of the ten correlated flares, AD~Leo event~C, AU~Mic event~C, AU~Mic
event~G and UV~Cet event~B show the strongest correlation between
radio and X-ray behaviour, and must be the strongest candidates for a
Neupert effect.

A few examples of radio flares with no X-ray counterpart were
observed. One particularly strong example in the lightcurve of AD~Leo
was found to be highly circularly polarized, suggesting a coherent
emission process and allowing the local magnetic field to be estimated
based on the assumption of an electron-cyclotron maser as the source.
The magnetic field strength suggested is of the order of kilogauss,
which would then imply that the kinetic energy available to generate X-ray emission
is very low.

Several examples of X-ray flares with no radio counterparts were also
observed. Searches were made at high time resolution for evidence of
very rapid continuous radio flaring, which might provide a heating
mechanism, but evidence for this was not compelling.  The lack of
radio emission associated with some X-ray flares could be due to the
emission being mostly synchrotron radiation from high energy
electrons, which is emitted preferentially perpendicularly to the
magnetic field lines.

 The overall behaviour of the sources is complex, and while a
  convincing case can sometimes be built for causal connections
  between individual radio and X-ray events, there are also many cases
  where radio events are {\em not} related to any X-ray event, and
  vice versa.  This then opens the possibility of coincidentally
  simultaneous or near-simultaneous flares which are not in fact
  related. We suggest that, for a causal connection to be coinsidered
  likely, temporal coincindence alone cannot be a reliable indicator
  but must be combined with other tests, such as correlation of
  structure in the time-resolved flare lightcurves, and a clear
  demonstration that the X-ray-radio relationship is energetically
  plausible.

\begin{acknowledgements}

We thank the referee for providing insightful and careful comments
which have helped us to improve the manuscript. This work is based
on observations obtained with {\em XMM-Newton}, an ESA science mission
with instruments and contributions directly funded by ESA Member
States and the USA (NASA). The VLA is a facility of the National Radio
Astronomy Observatory, which is operated by Associated Universities,
Inc., under cooperative agreement with the National Science
Foundation. This work has been funded in part by the Swiss National Science
Foundation (grant 20-66875.01).

\end{acknowledgements}

\end{document}